\input psfig
\documentstyle[11pt]{article}
\def\be{\begin{equation}}
\def\ee{\end{equation}}
\def\la{\label}
\def\bea{\begin{eqnarray}}
\def\eea{\end{eqnarray}}
\def\non{\nonumber}
\def\ci{\cite}
\def\la{\label}
\def\bib{\bibitem}

\def\lesssim{{_ <\atop{^\sim}}}
\def\lm{\lambda}
\def\le{\left}
\def\ri{\right}

\def\gm{\gamma}
\def\al{\alpha}

\def\fr{\frac}

\def\raw{\rightarrow}

\begin{document}

\begin{flushright}
  hep-ph/9909459 \\
  IFUNAM-FT-9906 \\
\end{flushright}

\vspace{15mm}

\begin{center}
   {\Large \bf  General Scalar Fields  as Quintessence }
\end{center}

\vspace*{0.7cm}

\begin{center}
{\bf A. de la Macorra\footnote{e-mail: macorra@fenix.ifisicacu.unam.mx} and G. Piccinelli\footnote{e-mail: gabriela@astroscu.unam.mx}}
\end{center}

\vspace*{0.1cm}

\begin{center}
\begin{tabular}{c}
{\small $^{1}$Instituto de F\'{\i}sica, UNAM}\\
{\small Apdo. Postal 20-364,
01000  M\'exico
D.F., M\'exico}\\
{\small $^{2}$Centro Tecnol\'ogico ENEP Arag\'on, UNAM}\\
{\small Av. Rancho Seco s/n, Col. Impulsora, Ciudad Nezahualcoyotl, M\'exico}\\
\end{tabular}
\end{center}

\vspace{1 cm}

\begin{center}
{\bf ABSTRACT}
\end{center}
{\small
We study the cosmological evolution of scalar fields with arbitrary potentials in the presence of a barotropic fluid (matter or radiation) without making any assumption on which term dominates.  We determine what kind of potentials $V(\phi)$ permits  a quintessence interpretation of the scalar field $\phi$ and to obtain interesting cosmological results. We show that all model dependence is given in terms of $\lambda \equiv - V'/V$ only and we study all possible asymptotic limits: $\lambda$ approaching zero, a finite constant or infinity. We determine the equation of state dynamically for each case. For the first class of potentials, the scalar field quickly dominates the universe behaviour, with an inflationary equation of state allowing for a quintessence interpretation. The second case gives the extensively studied exponential potential.
While in the last case, when $\lm$ approaches infinity, if it does not  oscillate then the energy density   redshifts faster  than the barotropic fluid  but if $\lm$ oscillates  then the energy density redshift depends on the specific potential.  
}


\noindent
\rule[.1in]{14.5cm}{.002in}

\thispagestyle{empty}

\setcounter{page}{0}
\vfill\eject

Models with a cosmological constant term, intended as a constant vacuum contribution
or as a slowly decaying scalar field, have recently received considerable attention
for several reasons, both theoretical and observational.

From the theoretical point of view, we have to face the possible conflict between
the age of the universe in the standard Einstein--de Sitter model and the age of the
oldest stars, in globular clusters.
Estimates of the Hubble expansion parameter from a variety of methods seem to point
to $H_0 \approx 70 \pm 10$ km/s/Mpc (a recent review can be found in  \ci{Freed};
see e.g. \ci{Ferr} for specific projects), leading to an expansion age of $t_U
\approx 9 \pm 1$ Gyr for a spatially flat universe with null cosmological constant.
On another hand, the age of globular clusters have been estimated in the range
$\simeq 13 \ - \ 15$ Gyr \ci{Chab0};
although revised determinations based on the Hypparcos distance scale are lower by
approximately 2 Gyr \ci{Chab}.

The requirements of structure formation models also suggest a cosmological constant
term. Simulations of structure formation profit from the presence of matter that
resists gravitational collapse and ${\rm \Lambda}$ CDM models provide a better fit to
the observed power spectrum of galaxy clustering then does the standard CDM model  \ci{EMS}, \ci{klypin}.

On the observational side, we find direct evidence in recent works on spectral and
photometric observations on type Ia supernova \ci{Riess} that favour
eternally expanding models with positive cosmological constant. Statistical fits to
several independent astrophysical constraints support these results \ci{Efst}. See
however \ci{Blinn} for a different explanation to these observations. More indirect
evidence comes from the observational support for a low matter density universe from
X--ray mass estimations in clusters \ci{White}, \ci{Steig}.
 In these works, if the nucleosynthesis limits on the
baryonic mass are to be respected, the total matter that clusters gravitationally is
limited to $\lesssim 0.3$. 
In such a case, a cosmological term would reconcile the low dynamical estimates of
the mean mass density with total critical density suggested by inflation and the
flatness problem.

Many models with a scalar field playing the role of a decaying cosmological constant
have been proposed up to now. Some of them are specific models motivated by physical
considerations but most of them are phenomenological proposals for the desired
energy density redshift \ci{RatP}-\ci{cc+redsh}. 
As a first step in the study of these models, the age of the Universe is calculated
for several redshift laws of the energy density that resides in the dynamical
scalar field \ci{Olson}, \ci{PeebR}.
Observational consequences of an evolving $\Lambda$ component decaying to matter
and/or radiation have been studied in \ci{Freese}, \ci{Wet},
obtaining severe constraints on such models. Another possibility, as the one that we
consider here, is that the scalar field couples to matter only through gravitation.
 This kind of scalar field , with negative pressure and a time-varying,
spatially fluctuating energy density, received the name of quintessence \ci{Cald}. Its
effects on the cosmic microwave background anisotropy are analysed in \ci{Cald} and
\ci{Silv} and phenomenological difficulties of quintessence have been studied in \ci{quint}. Constraints on the equation of state of a
quintessence--like component have been placed from observational data 
\ci{Turn}. Recently, a potential for a cosmologically successful decaying $\Lambda$ term has been constructed in \ci{Albr}.

As we have discussed above, the behaviour of scalar fields is fundamental in understanding the evolution of the universe. In this paper we are interested in  giving  a general approach to the analysis of the cosmological evolution of scalar fields and to determine what kind of potentials lead to a possible interpretation of the scalar field as quintessence and to a dominating
energy density.   However, we will not assume any kind of scale dependence for the potential   nor impose any condition on which energy density dominates \ci{gral}, \ci{cc+redsh},\ci{Wet},\ci{Vexp}-\ci{liddle}. We will show that all model dependence is given only in terms of the quantity $\lm\equiv -V'/V$, where the prime denotes derivative
with respect to the scalar field $\phi$, and its limiting behaviour at late times determines the evolution of the scalar field.
  
Our starting point is a universe filled with a barotropic  energy density, which can be either matter or radiation, and the energy density of a scalar field. The scalar field $\phi$ will have a self-interaction, given in terms of the scalar potential $V(\phi)$, but it will interact with all other fields only gravitationally. The barotropic fluid is described by an energy density $\rho_{\gm}$ and a pression $p_{\gm}$ with a  standard equation of state  $p_{\gm}=(\gm_{\gm}-1)\rho_{\gm}$, where $\gm_{\gm}=1$ for matter and $\gm_{\gm}=4/3$ for radiation. We do not make any hypothesis on which energy density dominates, that of the barotropic fluid or that of the scalar field.

The equations to be solved, for a spatially flat Friedmann--Robertson--Walker (FRW) Universe, are then given by
\bea
{\dot H}&=&-\frac{1}{2}(\rho_{\gm}+p_{\gm}+{\dot \phi}^2)\non \\
{\dot \rho}&=& - 3 H (\rho + p)
\la{cosmo} \\
{\ddot \phi}&=& - 3 H \dot\phi  - \frac{dV(\phi)}{d\phi}, \non
\eea
where $H$ is the Hubble parameter, $V(\phi)$ is the scalar field potential, $\dot{\phi}\equiv d\phi/dt$, $\rho \; (p)$ is the total energy density (pression) and we have taken $8 \pi G = 1$. It is useful to make a change of variables \ci{liddle} $x \equiv  {\dot \phi / \sqrt 6 H}$, $y \equiv  {\sqrt V / \sqrt 3 H}$ and eqs.(\ref{cosmo}) become
\bea
x_N&=& -3 x + \sqrt {3 \over 2} \lambda\,  y^2 + {3 \over 2} x [2x^2 + \gm_{\gm} (1 - x^2 - y^2)]  \non \\
y_N&=& - \sqrt {3 \over 2} \lambda \, x\, y + {3 \over 2} y [2x^2 + \gm_{\gm} (1 - x^2 - y^2)]
 \la{cosmo1} \\
H_N&=& -{3 \over 2} H [\gm_{\gm} (1-x^2-y^2) + 2x^2] \non
\eea
where $N$ is the logarithm of the scale factor $a$, $N \equiv ln (a)$, $f_N\equiv df/dN$ for $f=x,y,H$ and $\lambda (N) \equiv - V' / V$. Notice that all model dependence in eqs.(\ref{cosmo1}) is through the quantities $\lm (N)$ and the constant parameter $\gm_{\gm}$. Eqs.(\ref{cosmo1}) must be supplemented by the Friedmann or constraint equation for a flat universe
${\rho_{\gm} \over 3 H^2} + x^2 + y^2 = 1 $ and they
are valid for any scalar potential as long as the interaction between the scalar field and matter or radiation is  gravitational only. This means that it is possible to separate the  energy  and pression densities  into contributions from each component, i.e. $\rho=\rho_{\gm}+\rho_{\phi}$ and $p=p_{\gm}+p_{\phi}$, where   $\rho_{\phi}$ ($p_{\phi}$) is the energy density (pression) of the scalar field. We do not assume any equation of state for the scalar field. This is indeed necessary since one cannot fix the equation of state and the potential independently. For arbitrary potentials the equation of state for the scalar field  $p_{\phi}=( \gm_{\phi} -1)\rho_{\phi}$ is determined once $\rho_\phi, p_\phi$ have been obtained. Alternatively we can solve for $x, y$  using eqs.(\ref{cosmo1}) and  the quantity $\gm_{\phi}=(\rho_{\phi}+p_{\phi})/\rho_{\phi}=2x^2/(x^2+y^2)$ is, in general, time or scale dependent.

As a result of the dynamics, the scalar field will evolve to its minimum and if we do not wish to introduce any kind of unnatural constant or fine tuning problem, the minimum of the potential must have zero energy, i.e. $V|_{min}=V'|_{min}=0$ at $\phi_{min}$. We will consider here  only  these kind of potentials. For finite $\phi_{min}$ the scalar field will naturally oscillate around its vacuum expectation value (v.e.v.).  If the scalar field has a non zero  mass   or if the potential $V$  admits a Taylor expansion around $\phi_{min}$ than, using the H$\hat {\rm o}$pital rule, one has  lim$_{t \rightarrow \infty} |\lm| =\infty$ and it will oscillate. On the other hand, if $\phi_{min}=\infty$ then $\phi$ will not oscillate and  $\lm$ will approach either zero,  a finite constant  or  infinity. The oscillating behaviour of $\phi$ or $ \lm$ is important in determining the cosmological evolution  of $x, y$ and $ \Omega_{\phi}\equiv \rho_\phi/rho = x^2+y^2$ and we will show that  any scalar field with a non-vanishing mass  redshifts as  matter field.

Before solving eqs.(\ref{cosmo1}) we define the useful cosmological acceleration parameter $\alpha$ and expansion rate parameter $\Gamma$. The acceleration parameter is defined as
\be
\alpha\equiv \frac{\rho +3p}{(3\gm_{\gm}-2)\rho}=\frac{3\gm-2}{3\gm_{\gm}-2}
\la{al}
\ee
with $\gm =(\rho+p)/\rho$. 
If $\alpha=1$ then the acceleration  of the universe is the same as that of the barotropic fluid and any deviation of $\alpha$ from one implies a different cosmological behaviour of the universe due to the contribution of the  scalar field. A positive accelerating universe requires a negative $\alpha$ while for $0<\alpha<1$ the acceleration of the universe is negative (deceleration) but smaller than that of the barotropic fluid. For $\alpha>1$ the deceleration is larger than for the barotropic fluid.  In  terms of the standard deceleration parameter $q\equiv - \frac{\ddot{a} a}{\dot{a}^2}$ one has $\al=\frac{2q}{3\gm_{\gm-}2}$ or in terms of $x, y$  one finds $\alpha=1-\fr{3\gm_{\gm}}{3\gm_{\gm}-2}(y^2-x^2\frac{2-\gm_{\gm}}{\gm_{\gm}})
=1-3\Omega_\phi \fr{\gm_{\gm}-\gm\phi}{3\gm_{\gm}-2} $. It is also
useful to define the normalized equation of state parameter
\be
\Gamma=\frac{\gm}{\gm_{\gm}}
\la{Ga}
\ee
which gives the relative expansion rate of the universe with respect to the barotropic fluid. A $\Gamma$ smaller than one means that the universe expands slower than the barotropic fluid and a $\Gamma$ larger than one says that the universe expands faster due to the contribution of the scalar field. In our case $\alpha$ and $\Gamma$ are not independent since $\Gamma=1-(1-\alpha)(3\gm_{\gm}-2)/3\gm_{\gm}=1- \Omega_\phi\fr{\gm_{\gm}-\gm_{\phi}}{\gm_{\gm}}$.

A general analysis of eqs.(\ref{cosmo1}) can be done by noting that, given the constant parameter $\gm_{\gm}$, all model  dependence is  through the quantity $\lm(N)$. For an arbitrary potential $V$ eqs.(\ref{cosmo}) or (\ref{cosmo1}) will be, in general,  non-linear and there will be no analytic solutions. We can, of course,  solve them numerically but  we need to do it for each particular case and  initial conditions separately. In order to have an understanding on  the evolution of the scalar field  we will study the asymptotic limit.  It is useful to distinguish the different limiting cases for  the cosmological relevant quantities $x,y$ and $\Omega_{\phi}=x^2+y^2$. $\Omega_{\phi}$ will either approach zero, one or a finite constant value. For $\Omega_{\phi} \raw 0$ the scalar field dilutes faster than ordinary matter or radiation and if $\Omega_{\phi} \raw 1$ then the scalar  dominates the energy density of the universe. When $0<\Omega_{\phi}\raw cte < 1$ then the scalar and barotropic energy density redshift at the same speed.   Which behaviour will $x, y, \Omega_{\phi}$ have, depends on $\lm$  and on $\gm_{\gm}$. We will separate the analysis of eqs.(\ref{cosmo1}) into three different behaviours of $|\lm|$ at late times. In the first case we consider $\lm$  a finite constant (or approaching one), $\lm=c$. Secondly we study the limit $\lm \raw 0$. In the third case, we take  $\lm \rightarrow \infty$, which is the  natural case if the v.e.v. of $\phi$ is finite but we can have the same limit for  $\phi\raw \infty$. We divide in this case the analysis into an oscillating  and a not oscillating $|\lm|\raw \infty$.

Eqs.({\ref{cosmo1}) admit 5 different critical solutions for $\lm$ constant \ci{liddle}. Here, we generalise these attractors for more complicated potentials that have a non--constant $\lambda(N)$.
If $x,y$ do not oscillate, since their value is constrained to $|x| \leq 1,\, |y| \leq 1$, this implies that at late times they will approach a constant value, given by the attractor solutions of eqs.(\ref{cosmo1}), and $x_N, y_N$ will vanish. Three of the 5 different critical solutions $(x_c=1,y_c=0)$, $(x_c=-1,y_c=0)$ and $(x_c=0,y_c=0)$ are unstable (extreme) critical points. However, if $\lambda \raw \infty$ then the critical point $(x_c=0,y_c=0)$ becomes the asymptotic (stable) limit. The other two, depend  on the value of $\lambda(N)$.

For $\lm^2>3\gm_{\gm}$ \ci{liddle} one finds the "critical"   values
\be
x_c=\sqrt{\fr{3}{2}}\;\fr{\gm_{\gm}}{\lm}, \ \ \ y_c=\sqrt{\frac{3(2-\gm_{\gm})\gm_{\gm}}{2\lm^2}}, \ \ \ \Omega_{\phi c}=\frac{3\gm_{\gm}}{\lm^2}, \ \ \  \gm_{\phi}=\gm_{\gm}
\la{atr1}
\ee
for the quantities $x, y$, $\Omega_{\phi}$ and an effective equation of state equivalent to that of the barotropic fluid (i.e. $\gm_\phi=(\rho_\phi+p_{\phi})/\rho_\phi =\gm_{\gm}$). In this limiting case the redshift of the barotropic fluid and the scalar field is the same.

On the other hand if $\lm^2<6$ then one obtains
\be
x_c=\fr{\lm}{\sqrt{6}}, \ \ \  y_c=\sqrt{1-\fr{\lm^2}{6}}, \ \ \ \Omega_{\phi c}=1, \ \ \  \gm_{\phi}=\fr{\lm^2}{3},
\la{atr2}
\ee
and the scalar energy density dominates the universe. If the scalar field has $\gm_\phi < \gm_{\gm}$ then the solutions in eq.(\ref{atr2}) are stable,  the redshift of the scalar field is slower
than that of the barotropic fluid. However, if $ \gm_{\phi}=\lm^2/3  >\gm_{\gm}$,   then the solution in eq.(\ref{atr2}) is unstable and the scalar field ends up into the regime of the solution given in eq.(\ref{atr1}).

If $\lm (N)$ is constant then eqs.(\ref{atr1}) or (\ref{atr2}) are indeed solutions to $x_N=y_N=H_N=0$ but if $\lm (N)$ is not  constant then the critical values in eqs.(\ref{atr1}) and (\ref{atr2}) solve $x_N=y_N=H_N=0$ only on single points, not an interval. This means that they are a solution to eqs.(\ref{cosmo1}) only as an asymptotic limit and $x_c(N), y_c(N), \lm(N)$ are functions of $N$.

Let us now start with our  first case, i.e. $\lm$ constant. If $\lm=-V'/V=c$, the scalar potential has an exponential form, $V=he^{-c \phi}$. This case has been extensively studied \ci{Wet},\ci{Vexp}-\ci{liddle} and one finds critical (i.e. constant) points for $x$ and $y$ at late times. The value of $x, y$ depends on the value of $\lm=c$ and their solutions is given by eqs.(\ref{atr1}) or (\ref{atr2}). Since this case has been amply documented in the literature  \ci{Wet},\ci{Vexp}-\ci{liddle}, we do not 
include its numerical analysis here. We show in fig.(2g) the behaviour 
of $x$ and $y$ for comparison with a double exponential potential ($V=he^{-c e^\phi}$) (see below). The cosmological parameters are $\al=\Gamma=1$ for $\lm=c>\sqrt{3\gm_{\gm}}$ and $\al=\fr{c^2-2}{3\gm_{\gm}-2}, \; \Gamma=c^2/3\gm_{\gm}$ for $\lm=c=\sqrt{6}$. Note that in the first case $\Omega_\phi$ is finite and even if it dominates the universe  the acceleration and expansion of the universe is the same  as for the barotropic fluid. On the other hand, for  $\lm=c=\sqrt{6}$ one has $\Omega_\phi=1$  and the acceleration parameter $\al$ is in general different than one, it is negative if $c^2 < 2$ (assuming $\gm_{\gm}>2/3 $ i.e. matter or radiation).  In this case we could have  interesting  quintessence models.

For more complicated potentials, $\lm$ is not a constant and its evolution determines that of $x$ and $y$. The evolution of the scalar field leads to non-linear equations and critical points may exist but analytic solutions are either more difficult or impossible to obtain. However,
the solutions given in eqs.(\ref{atr1}) and (\ref{atr2})  may give  a good approximation on the limiting behaviour of $x$ and $y$.

Let us now consider the second case, i.e.  $\lm \rightarrow 0$. In this limit we can eliminate in eq.(\ref{cosmo1}) the term proportional to $\lm$, and since $-3 < H_N/H<0$ for all values of $x,y$ and $\gm_{\gm}$, we have
\be
\frac{x_N}{x}= -(3+\frac{H_N}{H})<0, \hspace{2cm} \frac{y_N}{y}= -\frac{H_N}{H} >0.
\la{lm=0}
\ee
From eq.(\ref{lm=0}) we conclude that $x$ will approach its minimum value (i.e. $x\raw 0$) while $y$ will increase to its maximum value (i.e. $y \raw 1$).  In the asymptotic region with $|x| \ll 1, |\lm| \ll 1$ one can solve eqs.(\ref{cosmo1}) for $x, y, H$ giving
\be
x(N)=\frac{e^{-3 N}}{\sqrt{1-c\,e^{-3\gm_{\gm} N}}}\,,\;\;\;    y(N)=\fr{1}{\sqrt{1-c\,e^{-3\gm_{\gm} N}}},\;\;\;H(N)=d\sqrt{1-c\,e^{-3\gm_{\gm} N/2}}
\la{lm0}\ee
 with $c,d$ integration constants. These solutions show that in the asymptotic region
the scalar field dominates the energy density of the universe and the Hubble parameter goes to a constant value.  

In the limit $\lm \raw 0$, the  first derivative of the potential approaches zero faster than the potential itself and examples of this kind of behaviour are given by potentials of the form $V=V_0 \phi^{-n}, n >0$.  The scalar field will dominate the energy of the universe leading to a "true" non-vanishing cosmological constant at late times with $x \raw 0, y \raw 1, \Omega_{\phi} \raw 1$ and $\gm_\phi \raw 0$.  The analytic solution to eq.(\ref{cosmo1}) for
$x, y$ can be approximated by the expressions given in eqs.(\ref{atr2}). However, $x, y$
are no longer constant since $x_c, y_c$ depend on $\lm(N)$, which is itself not constant but these expressions are a good approximation at late times (see fig.1d) to the numerical results.

The cosmological parameters are in this case  $\al=-\fr{2}{3\gm_{\gm}-2}$ and $\Gamma=0$. This means that 
the acceleration of the universe is positive (since $\al<0$) and the expansion of the universe is exponential. In fig.(1) we show the behaviour of  $x,y,\gm_{\phi}$ and $\Omega_{\phi}$ for a potential $V=V_0 \phi^{-1}$. In fig.(1a) we can see that the scalar field quickly dominates the universe
behaviour and the Hubble parameter tends to a constant different than zero, i.e. the universe
enters in an accelerated regime. This can also be seen in fig.(1b) where the
acceleration parameter $\alpha$ is smaller than $0$ and
$\gamma_\phi$, $\Gamma$ are inflationary almost all the time. 
Fig.(1c) shows the behaviour of $\lambda(N)$ for this case an din fig.(1d) we can see that the numerical solution has an asymptotic limit the solutions of eqs.(\ref{atr2}).
 
 \begin{figure}
\psfig{file=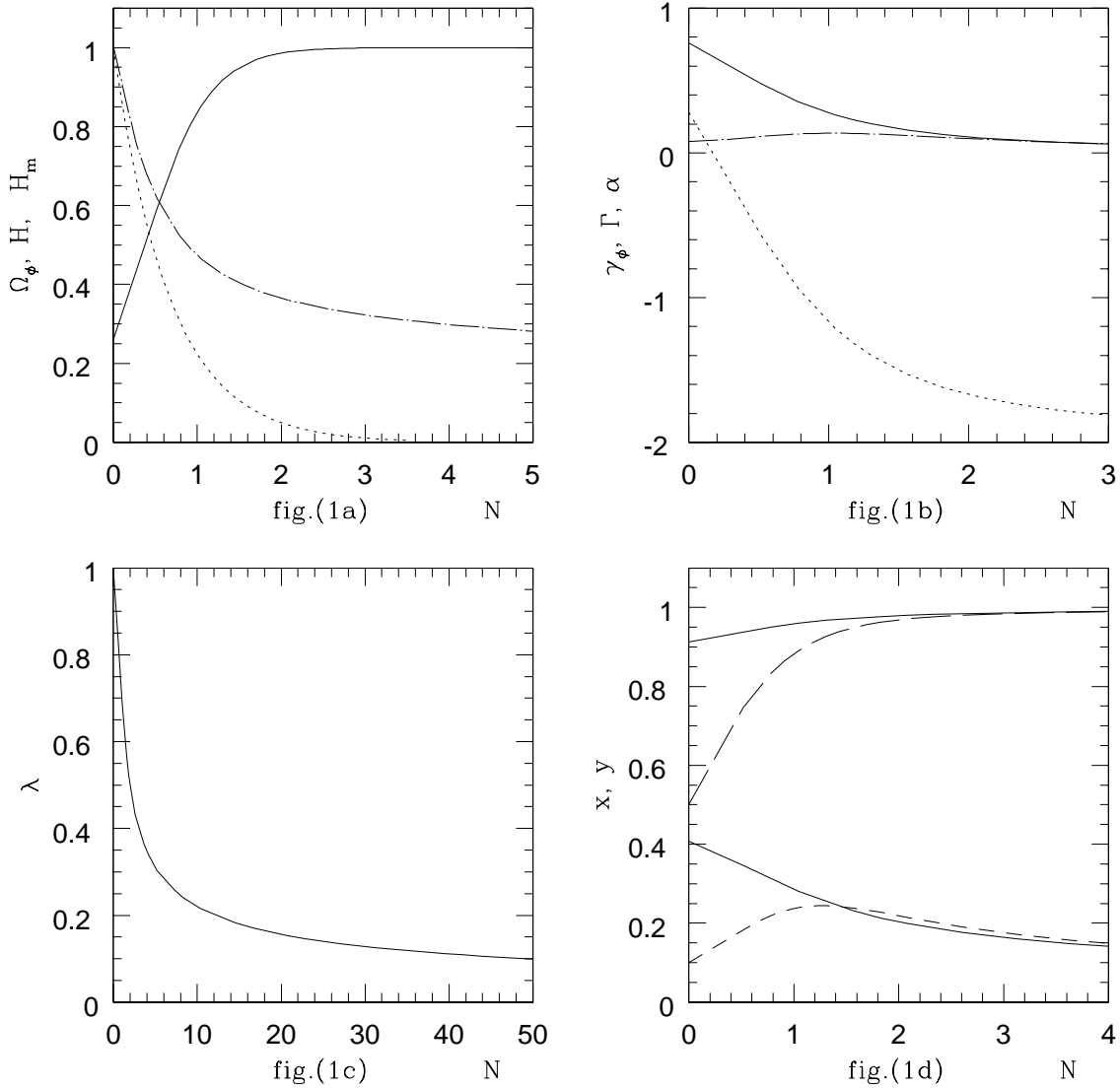,width=16cm,height=16cm}
\caption{\footnotesize{
 Evolution of the universe filled with matter and  a scalar field with $V = V_0 \phi^{-1}$. The initial conditions are $x_0=0.1$, $y_0=0.5$ and $H_0=1$.
In (1a) we show $\Omega_\phi$  (solid curve)  quickly approaching 1; the Hubble
parameter (dot--long dashed curve) tends to a finite
constant, and for comparison, we have drawn $H_{m}$ for a standard matter dominated universe
(dotted line); notice that with this type of potential the difference in the rate of
expansion with the standard model is remarkable. In (1b) we plot $\gamma_\phi$ (dot--long dashed curve), $\Gamma$ (solid curve) and
acceleration parameter $\alpha$ (dotted curve). In (1c) $\lambda$ slowly evolves to zero with $N$.
In (1d), the numerical $x$ and $y$ solutions are plotted (short and long
dashed respectively) and compared to their attracting solutions eq.(\ref{atr2}), lower
 and upper solid lines respectively.}}
 \end{figure}

As our final case we take the limit $\lm \raw \infty$, and we will separate this case into two different possibilities. The first one is when $\lm$ approaches its limiting value without oscillating and the later case is when $\lm$ does oscillate.

In the non--oscillating case, in the region $|\lm| \gg 1$ the leading term of eqs.(\ref{cosmo1})  is the one proportional to $\lm$ if $|y|, |x|$ are not much smaller than one. In such a case the eqs. for $x_N, y_N$ are
\be
x_N=\sqrt{\fr{3}{2}}\, y^2\, \lm, \hspace{2cm} y_N=-\sqrt{\fr{3}{2}} \,x\,y \,\lm.
\la{x'y'}\ee
The sign of $x_N$ is given by $\lm$ and if it does not  oscillate than $x$ would reach its maximum value $x=1$ while $y \raw 0$ for $\lm >0$ and $x \raw -1, y \raw 0$ for $\lm < 0$. However, in the region $y \raw 0$ the other terms in eq.(\ref{cosmo1}) become  relevant and eq.(\ref{x'y'})  is no longer a good approximation. In the region $|x| \ge |\lm| \,y^2$  the evolution of $x$ is given by  $x_N/x=-(3+H_N/H) < 0$ and $x$ will approach its minimum absolute value $x\raw 0$ like $y$. This region is the scaling region characterized by  almost constant values of $x,y$ and $\lm$. The end of the scaling region is when $y_N/y$ changes sign and becomes positive. This happens at $\sqrt{3/2} \lm x  + H_N/H \simeq 0$ with $H_N/H \simeq -3\gm_{\gm}/2$. After a brief increase for $y$ and $x$, they finally  end up approaching the values  given by the solution of eq.(\ref{atr1}) $x\raw x_c, y\raw y_c$ and going to the extreme values of $x=y=0$. In this case $x_c, y_c$ are not really critical (constant) points since $\lm$ is not constant.   The  kinetic and potential scalar energy will decrease faster than $\rho$, i.e. than ordinary matter or radiation. It is interesting to note that even though $x, y$ approach zero, the equation of state of the scalar field becomes constant, i.e. $\gm_\phi=2x^2/(x^2+y^2)=cte$ because $x,y$ decrease at the same velocity. This leads to  $\gm_{\phi}$ approaching the value of $\gm_{\gm}$, i.e. the equation of state of the scalar field will  be the same as that of the barotropic fluid (matter or radiation). Note in fig.(2h) that even though the equation of state of the scalar field approaches that of the barotropic fluid, its kinetic and potential energy decreases faster then that of the barotropic fluid, the reason being that $\gm_{\phi} \ge \gm_{\gm}$ at late times and the equality is only valid at $t=\infty.$ The cosmological parameters are $\al=\Gamma=1$ giving the same  asymptotic behaviour for the universe with or without the scalar field.
 
Examples of this kind of behaviour are given by potentials like $V=e^{-a \phi^2}, V=e^{-a e^{\phi}}$. In fig.(2) we show the behaviour of the dynamical variables and the cosmological
 parameters as a function of $N$ for $V=A e^{-c e^{\phi}}$ with 
$\lambda=-V'/V=c e^{\phi}$. This potential gives the asymptotic limit for string moduli fields \ci{mod}. The solution of eqs.(\ref{cosmo1}) shows that 
$\phi \raw \infty$ minimises the potential and $\lambda \raw \infty$ at late times
(fig.(2d)). The limiting values are $x=y=\Omega_{\phi}=0$, as we can see in 
fig.(2a)  and fig.(2h). In fig.(2a) we also show the evolution of the Hubble parameter in
our model, as compared with the standard matter dominated case; we can see that the
scalar field can influence the universe development only at early times. For small
$N$, the quantities $x,y$ have a similar evolution as for a simple exponential
potential.
We exhibit these quantities in fig.(2g) for a typical case $\gamma_\gamma = 1$ and
$\lambda = 10$. On the other hand, for $N \gg 1$, $x$ and $y$ are not constants and
they approach the value given in eqs.(\ref{atr1}), ($\lm^2$ is  larger than
$3\gm_{\gm}$), see fig.(2h). Fig.(2b) shows the behaviour of the $\gamma_\phi$
parameter as $x$ and $y$ evolve and the effective total $\Gamma$ parameter
(c.f. eq.(\ref{Ga})) for the ``fluid" composed by matter and
the scalar field. Fig.(2c) is the acceleration parameter defined in eq.(\ref{al})
Finally, fig.(2e) represents the phase space structure for $(x,y)$ obtained with
different initial conditions, where the final behaviour is amplified in fig.(2f).
The plateau in the graph for $\lambda$ in fig.(2d), for approximately 20 e--folds,
corresponds to the scaling region, where $x$ and $y$ are constants (almost zero in
this case), preceding the final evolution where the scalar field recovers a small
quantity of kinetic and potential energy that finally go to zero.

We conclude  that such fields are {\it not} good candidates for quintessence (parameterizing a slow varying cosmological constant) and they do not play a significant role at late times unless  they are produced at a late stage.  We would like to emphasise that these results are completely general and leave out a great number of candidate fields such as string moduli \ci{mod}. The only condition we have used to derive these results is that $\lm \raw \infty$ without  oscillating.

\begin{figure}
\begin{center}
\psfig{file=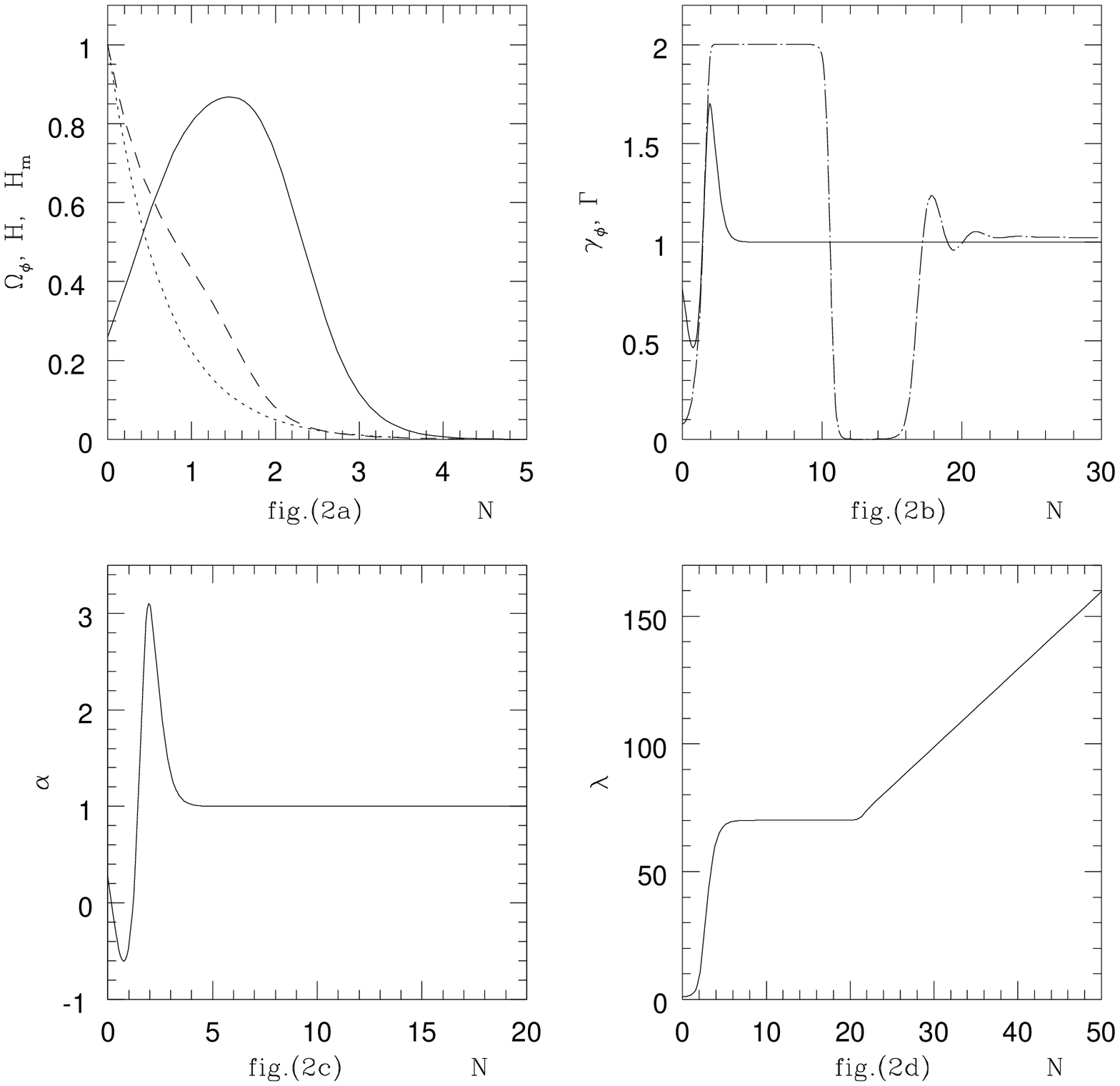,width=13cm,height=9cm}
\psfig{file=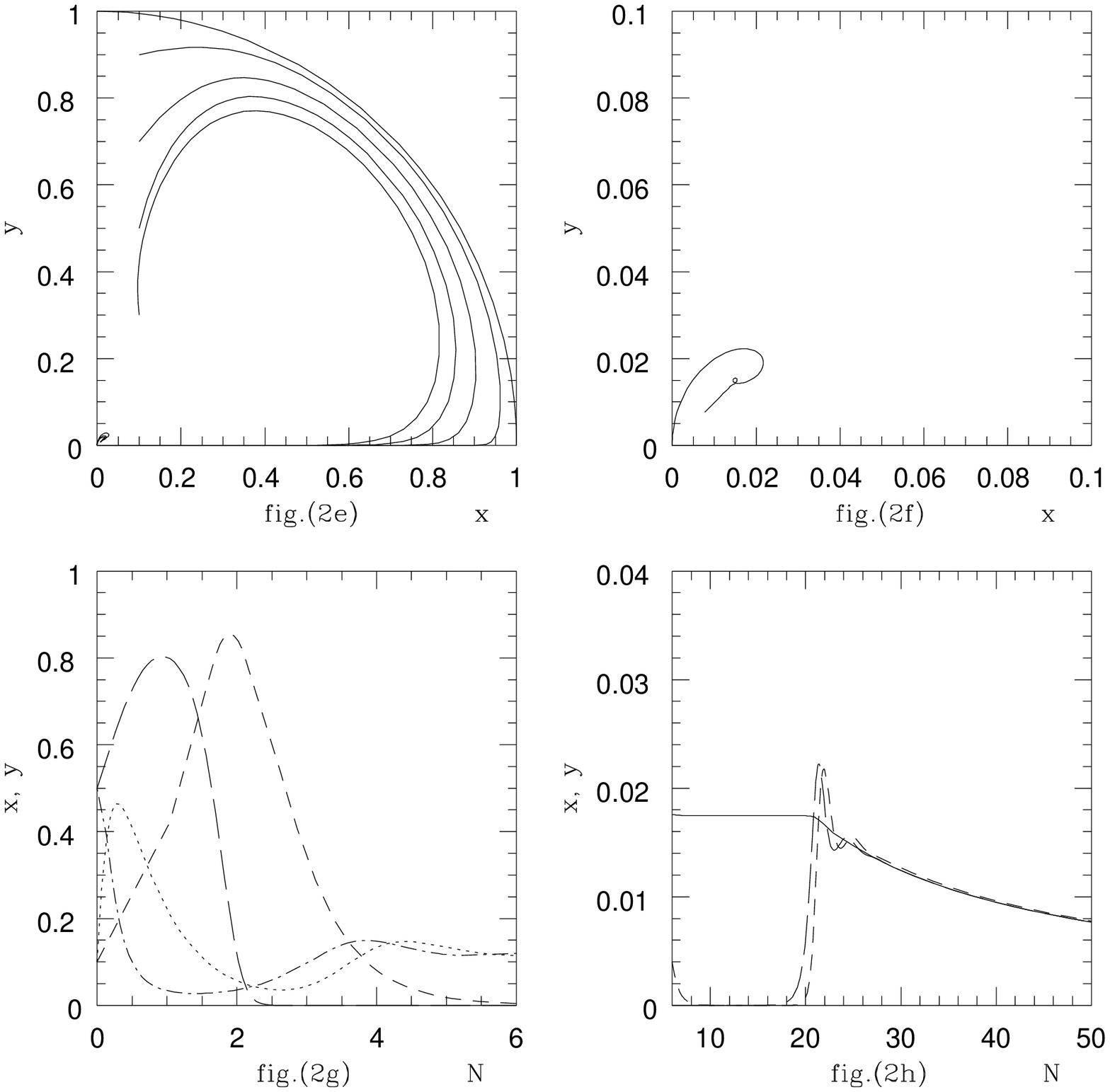,width=13cm,height=9cm}
\caption{\footnotesize{
Cosmological solution for $V=A~e^{-c e^{\phi}}$,
$\gamma_\gamma=1$ with initial
conditions  $x_0=0.1$, $y_0=0.5$, $\lambda_0=1$ and $H_0=1$. In
 (2a) the solid curve shows the evolution of $\Omega_\phi$, the dashed line is the
numerical solution for $H$, while the dotted line is the comparison
value of $H_{m}$ for a standard matter dominated universe, as a function of $N$.
In (2b) we show the evolution  of  $\gamma_\phi$   (dot-long dashed curve)
and of the effective total $\Gamma$ parameter (solid curve). In
(2c) the acceleration parameter $\alpha$ is displayed as a function of $N$. In
(2d) we have  $\lambda(N)$. We show in
(2e) the phase plane $(x,y)$ for different initial conditions $(x_0,y_0)$: 
$(0.0,1.0)$,$(0.1,0.9)$, $(0.1,0.7)$, $(0.1,0.5)$ and $(0.1,0.3)$; and the final
evolution is amplified in (2f).
(2g) compares the $x$ and $y$ evolution in this model (short and long--dashed curves
respectively) with the solution obtained in the purely exponential potential $V=h e^{-c\phi}$ case (dotted and dot--dashed curves respectively), with $\lambda=c=10$, for small $N$.
Finally, (2h) displays the asymptotic behaviour of $x$ and $y$ (curves are marked in
the same way as in (2g)), showing the good agreement with the attracting solution  eq.(\ref{atr1}) (solid curve).}}
\end{center}
\end{figure}

We will now consider the case when $|\lm| \raw \infty$ but with an oscillating $\phi$ field. In this case the v.e.v. of $\phi$ is finite and without loss of generality we can take it to be zero. Around the minimum the potential can be expressed as a power series in $\phi$ and keeping only the leading term we have $V=V_0 \; \phi^n$ with $n>0$ and even since the potential must be bounded. The  condition that $V'|_{min}=0$ requires $n>1$ and a finite scalar mass requires $n=2$.
For this potential $\lm=-n/\phi$ and it oscillates approaching a value $|\lm|=\infty$, see fig.(3). As a first guess we could think that the limiting behaviour of $x,y$ is also given by eq.(\ref{atr1}) and therefore tend to zero as $\lm \raw \infty$. However, this  asymptotic behaviour is no longer a good approximation and we must solve the dynamical non-linear equations (see fig.(3)
for the  numerical solution of a quadratic potential).

We  will now determine  under which conditions  $\Omega_{\phi}$   will either   dominate (approach one), oscillate around a finite constant value   or vanish. Since asymptotically $H \propto 1/t$,  a finite $\Omega_{\phi}$ ($\not= 1$) requires that $\dot\phi, (\phi)^{n/2} \propto 1/t$ or equivalently that $x,y$  are either constant or oscillate. We can thus write  $y=\frac{\phi^{n/2}}{3H}=B\; F_1^{n/2}[G(t)], \;  x=\frac{\dot\phi}{6H}=A\; F_2[G(t)]$ where $F_1,F_2$ are arbitrary oscillating functions  depending  on a single argument $G(t)$, and $A,B$ are constants. The function $G(t)$ is for the time being an unspecified function of $t$.  Of course, $F_1$ and $F_2$ are not independent since  the functional dependence of $\phi$ determines the functional dependence of $\dot\phi$, however this  is not important at this stage. Taking the derivative w.r.t. $N$,   we have $y_N=\frac{n F_{1G}}{2 F_1}\; G_N\,y\;$ and $x_N=\frac{F_{2G}}{F_2}\; G_N\, x$, where $ F_{iG}\equiv dF_i/dG, f_N\equiv df/dN$ with $f=x,y,G$ and $i=1,2$. Since $F_1,F_2$ are oscillating functions with a single argument $G(t)$, we have that the average of $<F_i^2>=<F_{iG}^2>$ and the average of $y^2,x^2$ is then
\be
<y_N^2>=\frac{n^2}{4}<G_N^2 y^2>, \hspace{2cm} <x_N^2>=<G_N^2 x^2>.
\la{avxy}
\ee
In the asymptotic limit with $x^2, y^2$ oscillating and $\lm \raw \infty$ the evolution of $x,y$ is given by eq.(\ref{x'y'}). Using eqs.(\ref{avxy}) and (\ref{x'y'}) we find that a potential $V=V_0\;\phi^n$ may have  finite values of $x,y$  at late times and
\be
\frac{<y^2>}{<x^2>}=\frac{2}{n},
\la{y/x}\ee
giving $<\gm_{\phi}>=2/(1+\frac{<y^2>}{<x^2>})=2n/(2+n)$. Depending if $\gm_{\gm}$ is larger, equal or smaller than $\gm_\phi$, $\Omega_\phi$ will go to one, finite constant or zero respectively. 
Notice that this result is completely general and  any massive scalar field redshifts at late times as matter fields with $\gm_\phi=\gm_{\gm}=1$ (i.e. $n=2$).

In order to obtain the asymptotic solution analytically, we will solve the  eqs.(\ref{cosmo1}) in a region where only one component of the energy density dominates. This will be  valid always in the asymptotic regime. However, we do not make any assumptions on which term dominates.
If it is the barotropic fluid which dominates then  the equation of state for $\rho$ will have a parameter $\gm=\gm_{\gm}$, however  if it is the scalar field that dominates than we take $\gm=<\gm_{\phi}>$ (average in time) since  in this case $\gm_{\phi}$ oscillates. In this regime we can solve for the Hubble parameter and we get the standard form $H(t)=2/(3\gm \,t)$. To determine the evolution of the scalar field we have to solve its equation of motion
${\ddot \phi} + 3 H \dot\phi  + \frac{dV(\phi)}{d\phi}=0$ with $V=V_0\; \phi^n$ and $n$ positive and even, i.e.
\be
{\ddot \phi} + \frac{2}{\gm \,t} \dot\phi  + n V_0 \phi^{n-1}=0.
\la{phi}
\ee
For $n=2$ the solution to eq.(\ref{phi}) is given in terms of $J_m$ and $Y_m$ the Bessel functions of the first and second kind respectively $\phi(z)=z^{-m}(2V_0)^{m/2}( c_1 J_m(z) + c_2 Y_m(z) )$ and \linebreak
$\dot\phi =-z^{-m}(2V_0)^{(m-1)/2}( c_1 J_{m+1}(z) + c_2 Y_{m+1}(z))$ with $c_1, c_2$ constants, $m\equiv (\gm-1/2)$,  $z\equiv t \sqrt{2V_0} $ and we have used that $d(z^{-m}K_m(z))/dz=-z^mK_{m+1}$  with $ K_m=J_m,Y_m$. Using these solutions we have
\bea
y&=& r z^{1-m} \le( k\, J_m(z) +  Y_m(z) \ri)  \non \\
x &=& -r  z^{1-m} \le( k\,J_{m+1}(z) +  Y_{m+1}(z) \ri) \la{xy2}
\eea
with  $r=\frac{3\gm}{8}\,c_2 (2V_0)^{\fr{m}{2}}, \;k\equiv c_2/c_1$. A simple analytic expression can be obtained
using the asymptotic limit of the Bessel functions  $J_m \simeq \sqrt{2/\pi z}\; cos(z-\frac{\pi(2m+1)}{4}), Y_m\simeq   \sqrt{2/\pi z} \;sin (z-\frac{\pi(2m+1)}{4})$ for $z \gg 1$ (i.e. $t \gg1)$. The amplitude of $x$ and $y$ in eq.(\ref{xy2}) in the limit $z\gg 1$ goes as $x\simeq  y \simeq z^{1/2-m}$. A finite value of $x,y$ requires $m=\gm-1/2=1/2$ (i.e. $\gm=1$). For $\gm > 1$ ($\gm < 1$) then $x,y \raw 1$ ($x,y \raw 0$) at large times. Furthermore, in  the asymptotic limit  $\gm_{\phi}=2( cos(z-\frac{\pi(2m+1)}{4})-k\, sin(z-\frac{\pi(2m+1)}{4}) )^2/(1+k^2)$ is an oscillating function with an average value $<\gm_{\phi}>=1$.  We can conclude therefore, that if the barotropic fluid has $\gm_{\gm} < 1$, i.e. smaller than $\gm_\phi$,  than $\Omega_{\phi}=x^2+y^2 \raw 0$, but if  $\gm_{\gm} > 1=<\gm_{\phi}>$ then the dominant energy density in the asymptotic regime will be that of the scalar field leading to $\Omega_{\phi}=1$. Finally, if $\gm_{\gm}=1=<\gm_{\phi}>$ then the energy density of the scalar field dilutes as fast as the barotropic fluid and $\Omega_\phi$ tends to a constant finite value.
The solutions in eq.(\ref{xy2}) for $\gm=1$ can be given  a completely analytic expression since in this case $m=1/2$ and the Bessel functions take simple form $J_{1/2}=\sqrt{2/\pi z}\; sin(z), Y_{1/2}=-\sqrt{2/\pi z}\;cos(z), J_{3/2}=\sqrt{2/\pi z}\;(sin(z)/z-cos(z))$ and $Y_{3/2}=-\sqrt{2/\pi z}\;(cos(z)/z+sin(z))$. Putting these expressions into the definitions of $x,y$  eq.(\ref{xy2}) we get
\bea
y&=&y_0 \;sin(z)-(x_0+\frac{y_0}{z_0}) cos(z), \la{xy2a} \non \\
x&=&-y_0 \;(\frac{sin(z)}{z}-cos(z)) +(x_0+\frac{y_0}{z_0})(\fr{cos(z)}{z}+sin(z))
\eea
where the initial conditions are given by $y_0, x_0$ at $z_0=\pi/2$.
In the limit $z \gg 1$ we have $\Omega_{\phi}\simeq y_0^2+(x_0+\frac{y_0}{z_0})^2$.

The analytic solution in eq.(\ref{xy2a}) agrees reasonably well  with the one obtained by solving eqs.(\ref{cosmo1}) numerically. This can be seen in fig.(3a), where we plot $\Omega_\phi$ obtained numerically for $\gamma_\gamma=1$ in a dot--long dashed line and expression (\ref{xy2a}) in a
solid line for a potential $V=V_0\;\phi^2$. We also plot $\Omega_\phi$ for $\gamma_\gamma = 4/3$ and $\gamma_\gamma= 1/2$ to illustrate the different asymptotic limits. 
For $\gm_{\gm} < 1$ we have $\Omega_{\phi} \raw 0$, $\gm_{\gm}=1$ gives $\Omega_{\phi} \raw cte$ and for $\gm_{\gm}=4/3$ we have $ \Omega_{\phi} \raw 1$. These different limits can be understood by noting that the average of $<\gm_{\phi}>=1$ and therefore, if $\gm_{\gm} > \gm_{\phi} (\gm_{\gm} < \gm_{\phi})$ the barotropic fluid  redshifts faster (slower) than the scalar field while for $\gm_{\gm}=<\gm_{\phi}>=1$ both energy densities dilute at the same speed.  The asymptotic value of $\Omega_\phi$ in the limiting case $\lm \raw \infty$ with an oscillating $\phi$ field depends  on the value of $\gm_{\gm}$ and on the initial conditions $x_0=x(N_0),y_0=y(N_0)$. Fig.(2b) shows the oscillating effective equations of state for the scalar
field and the mixture matter/scalar field, and the resulting acceleration features
of the universe $\alpha$.

 We have obtained the analytic solution of eq.(\ref{phi}) for $n=2$. This is clearly the simplest case since  eq.(\ref{phi}) is linear in $\phi$ and it's derivatives. For $n>2$ eq.(\ref{phi}) becomes non-linear in $\phi$ and no simple analytic solution exits. However, let us use in eq.(\ref{phi})  the ansatz
\be
\phi=t^{-2/n}\le( c_1 cos(\beta t^{2/n})+c_2 sin(\beta t^{2/n})\ri).
\la{defphi}\ee
It can be easily seen that this ansatz has the correct asymptotic
behaviour $\phi^{n/2}, \dot\phi, H \propto 1/t$ and $\ddot\phi \propto \phi^{n-1}$ if we wish to have $x,\,y $ finite. The $t$ exponents in eq.(\ref{defphi}) are determined by solving  eq.(\ref{phi}) and this equation  also imposes
the conditions $\gm=2n/(2+n)$ and $\beta=\fr{1}{4}V_0 n^3 c_2^{n-2}\;(sin(\beta t^{2/n})+k \;cos(\beta t^{2/n}))^{n-2}$. Notice that the value of $\gm$ is precisely
the value we  obtained using general arguments only (c.f. eq.(\ref{y/x})). Notice as well that the ansatz in eq.(\ref{defphi}) is not a ''complete" answer to eq.(\ref{phi}) since $\beta$ is not a true constant.  Only for $n=2$, $\beta$ is indeed constant and the solution in (\ref{defphi}) is the one we had previously obtained in terms of the Bessel functions (see eq.(\ref{xy2})). For $n\neq 2$ we must  take  the average of $\beta$ and work  in the asymptotic region. Nevertheless, eq.(\ref{defphi}) gives a good analytic approximation to the numerical solution. In terms of eq.(\ref{defphi}), $x,y$ take the following expressions,
\bea
x&=&\fr{x_0}{2/\pi + k}\le(k\;sin(\al t^{2/n})- cos(\al t^{2/n})+\frac{t^{-2/n}}{\al}(sin(\al t^{2/n})+k\; cos(\al t^{2/n}))\ri) \non \\
y&=&y_0\,\le(sin(\al t^{2/n})+k \;cos(\al t^{2/n})\ri)^{n/2}
\la{xyn}
\eea
with $ x_0=\fr{1}{n}\sqrt{3/2} \,\beta\, c_2\, \gm, \;  y_0=\fr{1}{2}\sqrt{3 V_0}\,c_2^{n/2}\,\gm$ and the initial conditions are taken at $ t_0^{2/n}=\pi/2\beta$. Using eq.(\ref{xyn}) we obtain $<y^2>/<x^2>=2/n$, at late times,  as in eq.(\ref{y/x}). We have, therefore, $<\gm_\phi>=2n/(2+n)$ and $\gm=<\gm_\phi>$, i.e. the ansatz in eq.(\ref{defphi}) is a "solution" to eq.(\ref{phi}) only when the dominant energy density redshifts as fast as the scalar field. This is, of course, no surprise since we imposed on the ansatz, eq.(\ref{defphi}), the limit $x,y \raw cte$.

 The cosmological parameters are in this case $\al=1-3\Omega_\phi\fr{\gm_{\gm}-\gm_\phi}{3\gm_{\gm}-2}$
and $\Gamma=1-\Omega_\phi\fr{\gm_{\gm}-\gm_{\phi}}{\gm_{\gm}}$ with $\gm_\phi=2n/(2+n)$.  From these expression we  conclude that when $\Omega_\phi$ remains finite $\al=\Gamma=1$ since in this case $\gm_{\gm}=\gm_\phi$ leading to the same behaviour of the universe with or without the contribution of the scalar field. However,  when $\Omega_\phi\raw 1$ then $\al=(3\gm_\phi-2)/(3\gm_{\gm}-2),\; \Gamma=\gm_{\phi}/\gm_{\gm}$ and one could have an accelerating universe  if  $\gm_\phi<2/3$  which requieres that $n<1$ and this is not acceptable since the first derivative of the scalar  potential must  vanish at the minimum. If $\Omega_\phi \raw 0$ then clealrly $\al=\Gamma=1$ and the scalar field plays asymptotically  no important role.

\begin{figure}
\begin{center}
\psfig{file=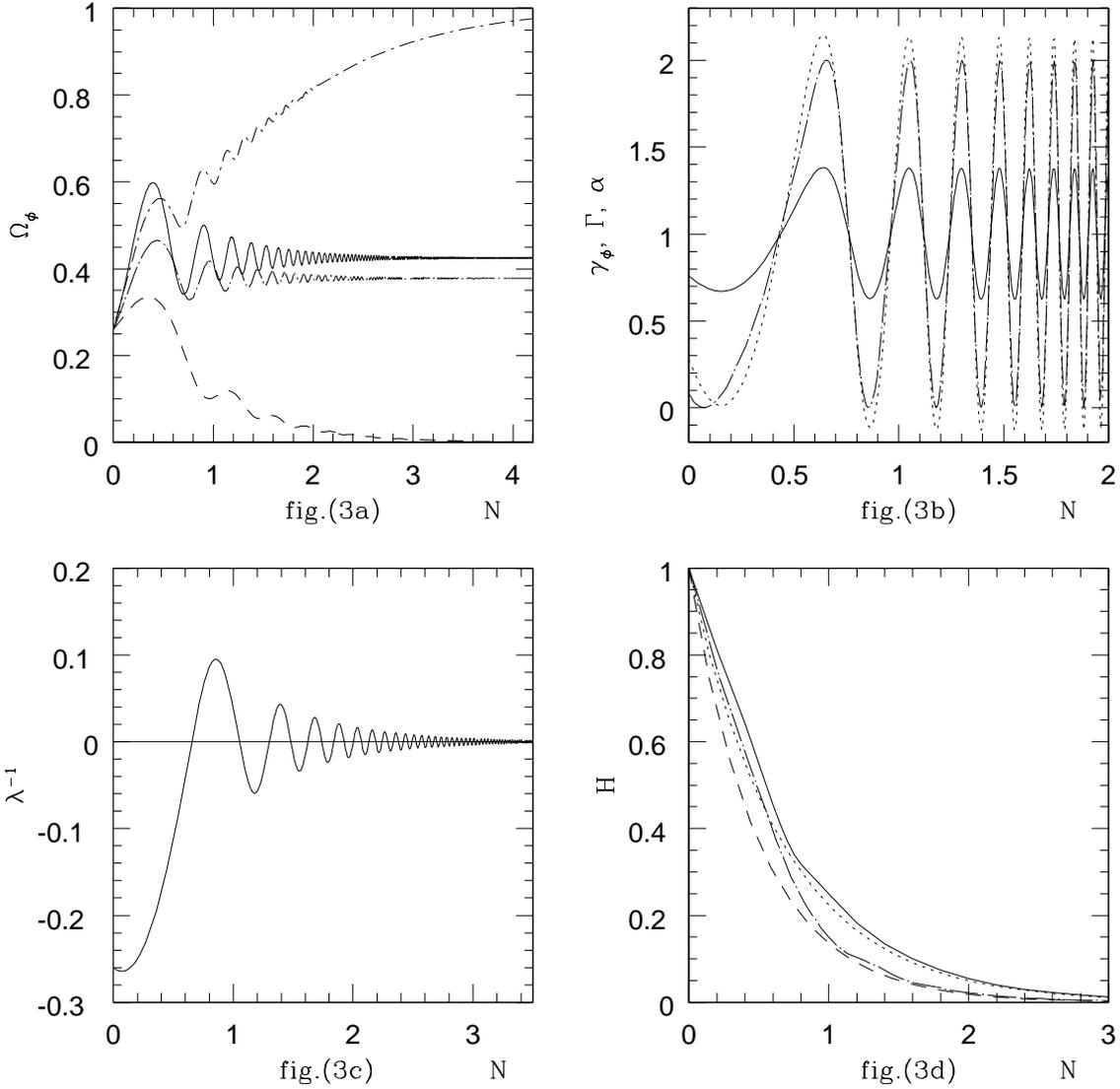,width=16cm,height=16cm}
\caption{\footnotesize{
Examples of evolution of the cosmological parameters for a universe filled
with a perfect fluid ($\gamma_\gamma = 1, 3/4, 1/2$) and a scalar field with
potential $V = V_0 \phi^2$. The initial conditions are
$x_0=0.1$, $y_0=0.5$, $H_0=1$ and $V_0=3\pi^2/32$. In fig.(3a) the numerical solution for $\Omega_\phi$ 
(dot--long dashed curve)  with $\gm_{\gm}=1$  is compared to the analytic expression $\Omega_\phi = x^2 +
y^2$, calculated from (\ref{xy2a}) (solid curve). Also shown in the figure are
$\Omega_\phi$ for $\gamma_\gamma = 3/4$ (dot--short dashed curve) and $\gamma_\gamma
= 1/2$ (dashed curve). In
(3b) the equation of state and acceleration parameters are displayed as a function
of $N$: $\gamma_\phi$ (dot--long dashed line), $\Gamma$ (solid line) and
acceleration parameter (dotted line). 
In (3c) we plot $\lambda ^{-1}$, in order to display the oscillating behaviour of
$\lambda (N)$, as it approaches $\infty$.
In (3d) we plot the evolution of $H$ for 2 different models. The solid curve
represents the numerical solution for $H$, with $V = V_0 \phi^2$ and $\gamma_\gamma
= 1$ and is compared to $H$ for a standard matter dominated universe (dotted curve).
The dot--long dashed curve corresponds to our numerical solution for $H$, with $V =
V_0 \phi^4$ and $\gamma_\gamma = 4/3$, following a similar evolution as a standard
radiation dominated model (dashed curve).}}
\end{center}
\end{figure}

To conclude this part of the analysis, we have established that if the initially dominant energy density component has  a $\gm$ parameter larger (smaller) than $<\gm_{\phi}>=2n/(2+n)$ then   $\Omega_{\phi}$ will
approach one (zero). For $n=2$ we have $<\gm_{\phi}>=1$,  for $n=4$ we have $<\gm_\phi>=4/3$.
Since the condition $V'|_{min}=0$ requires  $1<n$  we have  that $\gm_\phi >2/3$ far all $n$ and the scalar field  will not give an  accelerating universe. 
For  $n>4$ the energy density will decrease faster then radiation and since
at late times the universe is matter dominated, only  a scalar field with a non-vanishing mass  could lead to a significant contribution to the energy density of the universe. However, since its redshifts goes as matter   it is   not a candidate for  a cosmological constant  but it could serve as dark matter. 
Fig.(3d) illustrates these characteristics of a power-law potential for the Hubble parameter  $H$ for $V = V_0 \phi^n,\;n=2,4$, radiation and a matter dominated universe.

To summarise and conclude, we have studied the cosmological evolution of the universe filled with a barotropic fluid and a scalar field with an arbitrary potential but only with gravitational interaction with all other fields. The analysis done is completely general and 
we do not assume any kind of scale or time dependence of the scalar potential nor any assumption on which energy density  (barotropic or scalar) dominates. Our results are summarised in Table 1. We showed that all model dependence is given by $\lm\equiv -V'/V$ and $\gm_{\gm}$.  Any  scalar potential leads to one of the three different limiting cases of $\lm$:  finite constant, zero or infinity. In the first case, $\Omega_\phi$ approaches a finite constant (different than zero) depending  on the value of $\lm=c$.
For  $\lm \raw 0$ we obtained  $x\raw 0, y\raw 1$  with a  constant Hubble parameter $H$ and an accelerating  universe.
Finally,  for $\lm \raw \infty$ we concluded that if $\lm$ does not oscillate $x,y,\Omega_\phi \raw 0$ and if $\lm$ oscillates then all cases are possible (i.e. $\Omega_\phi \raw 0,1$ or a finite constant) depending on the value of $\gm_{\gm}$ and the power of the leading term in the scalar potential.  

\begin{center}
\begin{tabular}{|c|c|c|c|c|c|}
\hline
$\lm(\phi)=-V'/V$ & $\Omega_{\phi}=\rho_{\phi}/\rho$ & $\gm_{\phi}$ & $\alpha(\phi)$ & $\Gamma(\phi)$ & $e.g. V(\phi)$\\
\hline 
$c=$cte $(> \sqrt{3\gm_{\gm}} )$ &$ \frac{3\gm_{\gm}}{c^2} $&$ \gm_{\gm} $ & $1$ & $ 1$ & $  V_0\; e^{-c\phi}$\\
\hline
$c=$cte $( < \sqrt {6})$ &  1  &$  \frac{c^2}{3}$& $\fr{c^2-2}{3\gm_{\gm}-2}$ & $\fr{c^2}{3\gm_{\gm}}$ &  $ V_0\; e^{-c\phi} $\\
\hline
$\infty$ (no oscil.) & 0  & $ \gm_{\gm}  $ & $1$ & $1$ & $ V_0 \; e^{-c e^{\phi}}$\\
\hline
 & 0  &$ \frac{2n}{2+n} \;(> \gm_{\gm}) $& $ 1$ & $ 1$  \\
$\infty$ (oscil.) & cte  & $\frac{2n}{2+n} \;(= \gm_{\gm}) $& $ 1$ & $ 1 $ & $ V_0\; \phi^n,\; n>0$ even\\
  & 1  & $\frac{2n}{2+n} \;(< \gm_{\gm}) $&  $\fr{3\gm_\phi-2}{3\gm_{\gm}-2}$ & $ \fr{\gm_{\phi}}{\gm_{\gm}}$ &  \\
\hline
 0  &  1   &  0  & $-\fr{2}{3\gm_{\gm}-2}$ & $ 0$ &  $ V_0\;\phi^{-n}, \; n>0$\\
\hline
\end{tabular}
\end{center}
\footnotesize{ Table 1. In this table we show the asymptotic behaviour of $\Omega_{\phi},\; \gm_{\phi}$ the acceleration parameter $\al (\phi)=\frac{3\gm-2}{3\gm_{\gm}-2}$ and the expansion rate parameter $\Gamma=\frac{\gm}{\gm_{\gm}}$  for different limiting cases of $\lm (\phi)$. In the last column we give an example of  potential $V(\phi)$ which satisfies this limit.
 }}

G.P. would like to thank the Instituto de Astronom\'{\i}a, UNAM, for their kind 
hospitality during the realization of this work and partial support from DGAPA, UNAM, project IN-109896. A.M. research was supported in part by CONACYT
 project 32415-E  and by DGAPA, UNAM,  project IN-103997.

\thebibliography{}
\footnotesize{
\bibitem{Freed} W.L. Freedman, {\it David Schramm Memorial Volume}, Phys. Rep., in
press (2000),  astro-ph/9909076.
\bibitem{Ferr} {L. Ferrarese {\it et al.}, {\it Proceedings of the Cosmic Flows
Workshop} (1999), astro-ph/9909134 ; M.A. Hendry and S. Rauzy, {\it Proceedings of
the Cosmic Flows Workshop} (1999),  astro-ph/9908343.}
\bibitem{Chab0} B. Chaboyer, P. Demarque, P.J. Kernan and L.M. Krauss, Science 271 (1996)
957.
\bibitem{Chab} B. Chaboyer, P. Demarque, P.J. Kernan and L.M. Krauss, ApJ 494 (1998) 96.
\bibitem{EMS} G. Efstathiou, S. Maddox and W. Sutherland, Nature 348 (1990) 705.
\bibitem{klypin} {J. Primack and A. Klypin, Nucl. Phys. Proc. Suppl. 51 B, (1996),
30; J.F. Navarro and M. Steinmetz ,  astro-ph/9908119.}
\bibitem{Riess} {A.G. Riess {\it et al.}, Astron. J. 116 (1998) 1009; S.
Perlmutter {\it et al}, ApJ 517 (1999) 565; P.M. Garnavich {\it et al}, Ap.J 509 (1998)
74.}
\bibitem{Efst} {G. Efstathiou, S.L. Bridle, A.N. Lasenby, M.P. Hobson and R.S. Ellis,
MNRAS 303L (1999) 47; M. Roos and S.M.
Harun-or Rashid, astro-ph/9901234.}
\bibitem{Blinn} {E.I Sorokina, S.I. Blinnikov and O.S. Bartunov (astro-ph/9906494);
A.G. Riess, A.V. Filippenko, W. Li and B.P. Schmidt, 
astro-ph/9907038.}
\bibitem{White} S.D.M. White, J.F. Navarro, A.E. Evrard and C.S. Frenk, Nature 366 (1993)
429.
\bibitem{Steig} G. Steigman and J.E. Felten, {\it Proceedings of the St. Petersburg Gamow
Seminar}, ed. A.M Bykov and R.A. Chevalier, Sp. Sci. Rev. (1995)
\bibitem{RatP} {B. Ratra and P.J.E. Peebles, Phys. Rev. D37 (1988) 3406} 
\bib{gral}{J.M
Overduin and F.I. Cooperstock, Phys. Rev. D58 (1998) 20; A.R. Liddle and R.J. Scherrer,Phys. Rev. D59,  (1999)023509;
 V. Mendez, Class Quant. Grav. 13(1996) 3229;
 J. Uzan, Phys. Rev. D59, (1999) 123510; A.P. Billyard, A.A. Coley, R.J. van den Hoogen, J. Ibanez and I. Olasagaste, gr-qc/9907053}
\bib{cc+redsh}{W. Chen and Y. Wu, Phys. Rev D41 (1990) 695;  Y. Fujui and T. Nishioka, Phys. Rev. D42 (1990) 361; D. Pavon, Phys. Rev. D43 (1991) 375; J. Matyjasek, Phys. rev. D51 (1995) 4154; M. S. Berman,  Phys. Rev. D43 (1991) 1075; J.C. Carvalho, J.A.S. Lima and I. Waga, Phys. Rev. D46 (1992) 2404; J.A.S. Lima and J.M.F. Maia, Phys. Rev. D49 (1994) 5597; V. Silveira and I. Waga, Phys. Rev D50 (1994) 489} 
\bibitem{Olson} {T.S. Olson and T.F. Jordan, Phys. Rev. D 35 (1987) 3258; J. W. Moffat, Phys. Lett. B357 (1995) 526}
\bibitem{PeebR} {P.J.E. Peebles and B. Ratra, ApJ 325 (1988) L17; J.W. Moffat, Phys. Lett. B 357 (1995) 526.}
\bibitem{Freese} {K. Freese, F.C. Adams, J.A. Frieman and E. Mottola, Nucl. Phys. B 287
(1987) 797; M. Birkel and S. Sarkar, Astropart. Phys. 6 (1997) 197.} 
\bib{Wet}{C. Wetterich, Astron. Astrophys.301 (1995) 321
\bibitem{Cald} R.R. Caldwell, R. Dave and P.J. Steinhardt, Phys. Rev. Lett. 80 (1998) 1582.
\bibitem{Silv} {V. Silveira and I. Waga, Phys. Rev. D (1997) 4625; L. Amendola
astro-ph/9908023.}
\bib{quint}{D. Lyth and C. Koldo, Phys. Lett. B458 (1999) 197; I. Zlater, L. Wang and P.J. Steinhardt, Phys. Rev. Lett.82 (1999) 896; D. Huterer and M.S. Turner astro-ph/9808133; P. Brax and J. Martin, astro-ph/9905040; T. Chiba, gr-qc/9903094}
\bibitem{Turn} {M.S. Turner and M. White Phys. Rev. D 56 (1997) 4439; G. Efstathiou,  astro-ph/9904356; L. Wang {\it et al},
astro-ph/9901388.}
\bibitem{Albr} A. Albrecht and  C. Skordis, astro-ph/9908085. 
\bibitem{Wet2} C. Wetterich, Nucl. Phys. B302 (1998) 668 
\bibitem{Vexp} {P. Ferreira, M. Joyce, Phys. Rev. D 58 (1998) 503; E.J. Copeland, A. Liddle and D. Wands, Ann. N.Y. Acad. Sci. 688 (1993) 647.   }
\bib{liddle}E.J. Copeland, A. Liddle and D. Wands, Phys. Rev. D57 (1998) 4686
\bib{mod} A. de la Macorra in preparation
 
}

\end{document}